\documentclass[11pt]{article}

\usepackage{latexsym, amsmath, amssymb}
\usepackage{psfrag}
\usepackage{braket}
\usepackage{graphicx}
\usepackage{mathrsfs}
\usepackage{bbm}
\usepackage{psfig}

\usepackage{latexsym, amsmath, amssymb}
\usepackage{psfrag}
\usepackage{psfig}
\usepackage{subfigure}
\usepackage{graphicx}
\usepackage{mathrsfs}
\usepackage{bbm}

\newcommand{\half}{{1\over2}}
\newcommand{\be}{\begin{eqnarray*}}
\newcommand{\ee}{\end{eqnarray*}}
\newcommand{\bea}{\begin{eqnarray}}
\newcommand{\eea}{\end{eqnarray}}
\newcommand{\ms}{\medskip}

\newcommand{\bzA}{\bar z_{AdS}}
\newcommand{\bz}{\bar z}
\newcommand{\halfbps}{$\half$-BPS}
\newcommand{\adsS}{AdS$_5\times S^5$}
\newcommand{\p}{\partial}
\newcommand{\Title}[1]{
  ~\vspace{12mm}
  \begin{center}
    \Large\bf #1
  \end{center}
  }
\newcommand{\Author}[1]{
  \begin{center}
    \large #1
  \end{center}
  }
\newcommand{\Institution}[1]{
  ~\vspace{-26pt}
  \begin{center}
   #1
  \end{center}
  }
\newcommand{\Email}[1]{{\tt #1}}

\newcommand{\History}[1]{
\centerline{\small #1}
}

\begin{document} \setlength{\unitlength}{1mm}

\thispagestyle{empty}
\rightline{\tt hep-th/0507260}
\Title{Black Hole Statistics from Holography}
\bigskip
\Author{Peter G. Shepard}
\Institution{
\it{Center for Theoretical Physics and Department of Physics, University of California\\
Berkeley, CA 94720-7300, USA\\
and\\
Theoretical Physics Group, Lawrence Berkeley National 
Laboratory\\
Berkeley, CA 94720-8162, USA}}
\begin{center}
\Email{pgs@socrates.berkeley.edu}
\ms\ms
\History{July 27th, 2005}
\end{center}
\ms\ms

\begin{abstract} 
We study the microstates of the ``small'' black hole in the $\half$-BPS sector of AdS$_5\times S^5$, the superstar \cite{Myers}, using the powerful holographic description provided by LLM \cite{llm}.  The system demonstrates the inherently statistical nature of black holes, with the geometry of \cite{Myers} emerging only after averaging over an ensemble of geometries.  The individual microstate geometries differ in the highly non-trivial topology of a quantum foam at their core, and the entropy can be understood as a partition of $N$ units of flux among 5-cycles, as required by flux quantization.  While the system offers confirmation of the most controversial aspect of Mathur and Lunin's recent ``fuzzball'' proposal \cite{Lunin}\cite{Mathur}, we see signs of a discrepancy in interpreting its details.

\newpage

\end{abstract}
\section{Introduction}

By defining a full quantum-mechanical description of gravity in terms of a dual gauge theory, holography is expected to provide resolution to many outstanding questions in general relativity.  However, since it is often very difficult to construct both sides of the duality, many questions have been slow in revealing their mysteries.  A beautiful recent work by Lin, Lunin and Maldacena \cite{llm} has managed to explicitly solve both sides of a non-trivial sector of the AdS/CFT duality \cite{AdSCFT}, revealing fascinating new structure, and opening up the possibility to return to some of these perennially difficult issues.  Among other things, this has already been used to better understand chronology protection, by demonstrating its connection to the Pauli principle in the dual theory \cite{Klemm} \cite{Milanesi:2005tp}, and to study topology changing transitions, showing that this class of geometries has a fundamental transition similar to the conifold transition, which is a smooth merger of fermi droplets in the dual theory \cite{us}.  Since LLM provides perhaps the most transparent realization of quantum gravity to date, it seems that any mystery of quantum gravity should be tackled there first.  In this paper, we will use this elegant description of quantum gravity to attempt to better understand the black hole information paradox.  

As Mathur and Lunin have pointed out \cite{Lunin} \cite{Mathur}, the fact that holographic dualities are a one-to-one map between gauge theory and geometry implies that microscopic accounts of black hole entropy by a gauge theory, such as that of Strominger and Vafa \cite{Strominger}, should have a description entirely in terms of microstates of the geometry alone.  Mathur and Lunin have boldly conjectured that all black holes arise only upon coarse-graining over such microstate geometries.

   By studying the ``small'' black hole contained in the LLM sector, we will test these ideas.  Though we will encounter difficulty with some details of Mathur and Lunin's conjectures, we will find a vivid illustration of the inherently statistical-mechanical nature of the black hole.  We will see that quantum effects modify the naive classical geometry to account for the black hole's entropy, replacing a single singular metric with an ensemble of smooth geometries, with the singularity replaced by a quantum foam of highly non-trivial topology.

\section{The Superstar}
Though the \halfbps\ sector of excitations on \adsS\ does not contain a macroscopic black hole, there is a non-supersymmetric family of asymptotically \adsS\ black holes discovered by Myers and Tafjord \cite{Myers} whose singular extremal limit preserves half the supersymmetry.  The full family is governed by 5 parameters: the 5-form flux threading the $S^5$ $N$, 3 angular momenta on the $S^5$ $J_i \in$ SO(6), and an excess mass parameter $\mu$.  The extremal limit sets $\mu=J_1=J_2=0$, leaving a two-parameter family given by $N$ and $\Delta \equiv J_3$ called the superstar \cite{Myers}, \cite{Klemm}:
\bea
ds^2 &=& - \frac{1}{G}\left(\cos^2 \theta + {R^2 \over R_{AdS}^2}G^2\right)dt^2+R_{AdS}^2 {H \over G}\sin^2 \theta d\phi^2 \nonumber \\
&&+2{R_{AdS} \over G}\sin^2 \theta dt d\phi + G \left({dr^2 \over f}+R^2 d\Omega_3^2\right)\label{superds}\\
&&+R_{AdS}^2G d\theta^2+ {R_{AdS}^2 \over G}\cos ^2 \theta d \tilde \Omega_3^2. \nonumber
\eea
with 
\bea
f &=& 1+H({R\over R_{AdS}})^2 \label{f}\\
G &=& \sqrt{\sin^2 \theta + H \cos^2 \theta}\\
H &=& 1+2{R_{AdS}^2\over R^2}{\Delta \over N^2}\label{H}.
\eea

Though the superstar has a naked singularity, this geometry should be a valid testing ground of black hole physics since it satisfies Gubser's criterion for ``good'' naked singularities \cite{Gubser}.  According to his conjecture, since the superstar singularity can be thermalized to a legitimate black hole, it is expected that $\alpha'$ corrections to the equations of motion would yield a geometry with a finite-area horizon.  Indeed, Dabholkar \cite{Dabholkar} has used the attractor mechanism to exactly account for these corrections for singular 2-charge black holes, to demonstrate a naked singularity resolved to yield a finite-area horizon.  We will therefore assume that $\alpha'$ corrections would lift the superstar to a true black hole, and refer to the superstar as a black hole for the remainder of this paper.

As a \halfbps\ excitation of \adsS, the superstar is in the sector of AdS/CFT solved by LLM, allowing us a window into black hole physics in a full quantum-mechanical setting, described as free fermions in a harmonic oscillator potential.  

\section{Review of LLM}

Following work of Corley, Jevicki and Ramgoolam \cite{Jev} and of Berenstein \cite{Beren}, who solved the CFT, LLM were able to completely solve the $\half$-BPS sector of the AdS/CFT correspondence, as we will review very briefly here.

\subsection{The Field Theory}
Beginning with ${\cal N}=4$ SYM on $S^3\times{\bf R}$, the $\half$-BPS sector is given by picking out a generator $J$ in the SO(6) ${\cal R}$-symmetry algebra, say that which acts as a phase on the complex scalar $Z=\phi^1+i\phi^2$, and imposing $\Delta=J$.  This condition requires almost every field to be in its ground state, allowing only fluctuations of the adjoint-valued field $Z$ which are constant on the sphere.  As such, the field theory reduces to the quantum mechanics of the s-wave of $Z$, a gauged matrix model, with a harmonic oscillator potential term due to the conformal coupling of $Z$ to the Ricci scalar of $S^3$.  As discussed in \cite{Jev}, the BPS condition further reduces the excitations to those of a single {\it hermitian} matrix $\Phi$, with action
\be
S &=& \half \int dt\,\rm{Tr}\left\{(D_t\Phi)^2-\Phi^2\right\}.
\ee
By the usual arguments (see \cite{Kleb}, for example), a gauge choice for the auxiliary U(N) gauge field $A_0$ in the covariant derivative, $D_t=\partial_t-iA_0$, reduces the matrix $\Phi$ to its eigenvalues $q_i$, $i=1 \dots N$.\footnote{In the large-$N$ limit implicitly considered here, we may ignore the distinction between U(N) and SU(N), which is an ${\cal O}(1/N)$ correction.}  In the change of variables from the $N^2$ matrix elements to $N$ eigenvalues, there is a Jacobian factor of the antisymmetric Vandermonde determinant, which has the crucial effect of turning the eigenvalues into non-interacting fermions.  

\begin{figure}
        \centerline{\psfig{file=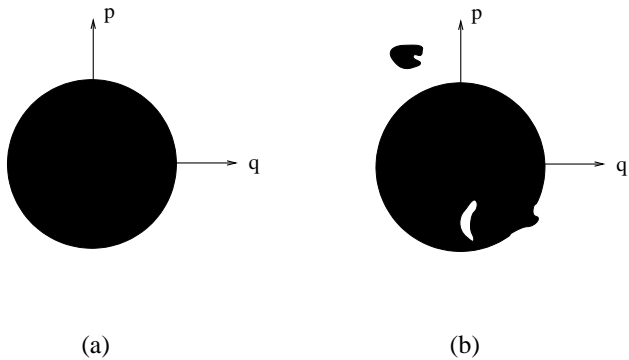,width=0.5\textwidth}}
        \caption{a) The ground state of the matrix model, $AdS_5\times {\bf S}^5$.  b) A hole, large excitation, and ripple of the fermi surface; giant, dual-giant and graviton, respectively.  The plane is alternately phase space $(p,q)$ in the CFT or the $y=0$ $(x_1,x_2)$ plane of the geometry.}
        \label{fig:blob}
\end{figure}

The BPS states of the CFT are therefore in one-to-one correspondence with the states of $N$ free fermions in a harmonic oscillator potential, whose state in a semi-classical phase-space $(q,p)$ is specified by droplets of a two-dimensional quantum Hall fluid (Figure \ref{fig:blob}).

\subsection{The Geometries}

By picking out a 2-plane in the $\cal{R}$-symmetry group SO(6), but preserving the isometries of $S^3 \times {\bf R}$, the above states generically break the bosonic symmetry of the system to $SO(4)\times SO(4)\times \bf{R}$ and break half the supersymmetry.  By imposing the Killing spinor equation for a metric ansatz with these isometries, LLM were able to solve for the most general solution to Type IIB supergravity with this symmetry.  The $SO(4)\times SO(4)$ isometries select out two privileged $S^3$'s, with radii $R$ and $\tilde R$, and the Hamiltonian ${\bf R}$ selects a privileged time coordinate $t$.  Additionally, manipulation of the spinor equations reveal the existence of an $8^{th}$ special coordinate $y$, with the property $R \tilde R= y$, making the plane $y=0$ very important.  

In terms of the two remaining coordinates, $x_1$ and $x_2$, the metric takes the form
\bea
ds^2=-h^{-2}(dt+V_idx^i)^2+h^2(dy^2+dx^idx^i)+R^2d\Omega_3^2
+\tilde R^2d\tilde\Omega_3^2. \label{eesolmet}
\eea
Remarkably, all of the coefficients in (\ref{eesolmet}) are determined in terms of a single function $z(x_1,x_2,y)$.  For example,
\bea
R^2 &=& y\sqrt{\frac{1+2z}{1-2z}},\nonumber\\
\tilde R^2 &=& y\sqrt{\frac{1-2z}{1+2z}},\label{radii}\\
h^{-2} &=& \frac{2y}{\sqrt{1-4z^2}}=R^2+\tilde R^2.\nonumber
\eea
See \cite{llm} for expressions for $V_i$ and for the 5-form field strength.  The dilaton, axion, and other Ramond-Ramond fields vanish.  The function $z$ obeys a simple second order linear equation in the space $(x_1,x_2,y)$,
\be
\p_i\p_iz+y\p_y\left(\frac{\p_yz}{y}\right) &=& 0,\label{eezeeeq} 
\ee
which determines the function in terms of the boundary data at $y=0$.  Since the product of the radii of the spheres vanishes on this plane, singularities are avoided only if $z(x_1,x_2,y=0)$ takes the values $\pm \half$ there, so that one or the other $S^3$ remains finite.  A solution is then fully specified by the data of regions of the $x_1$-$x_2$ plane on which $z(x_1,x_2,y=0)$ takes each value, which \cite{llm} color-code as $Black=-\half$, where $S^3$ shrinks, and $White=+\half$, where $\tilde S^3$ shrinks. 

To connect these geometries to their holographic dual, LLM identify this boundary data with the phase space of the fermions, $(x_1,x_2)=(q,p)$.  The non-singularity condition $\half-z(y=0)\equiv \rho\in \{0,1\}$ is understood as the fermion occupation number, while quantization of fermion number matches with 5-form flux quantization in the bulk.  Additionally, quantization of phase space to a minimal area $\hbar=2\pi l_p^4$ is an indication of a space-time non-commutativity \cite{llm} (see also \cite{Jab}).  For example, AdS$_5\times S^5$ is given by the solid disk, with radius $r_0^2=R_{AdS}^4=4\pi l_p^4 N$, (Figure~\ref{fig:blob}a), as it should to agree with the ground state of the CFT.  This yields the following metric functions in terms of polar coordinates $r,\ \phi$ on the $x_1$-$x_2$ plane:
\bea
z(r,y;r_0) &=& \frac{r^2- r_0^2+y^2}{2\sqrt{ (r^2+r_0^2+y^2)^2-4r^2r_0^2}} \label{zads}\\
V_\phi(r,y;r_0)&=&-\frac{1}{2} \left( \frac{r^2+y^2+r_0^2}{\sqrt{(r^2+ r_0^2+y^2)^2-4 r^2r_0^2}}-1 \right). \nonumber
\eea
By the change of coordinates
\be
y &=& r_0 \sinh \rho \sin \theta \\
r &=& r_0 \cosh\rho \cos \theta\\
\tilde \phi &=&  \phi - t
\ee
we get the standard form of the $AdS_5 \times {\bf S}^5$ metric
\be
ds^2 &=& r_0 [ - \cosh^2 \rho d t^2 + d\rho^2+ \sinh^2 \rho d\Omega_3^2 +
d \theta^2 + \cos^2 \theta d \tilde \phi^2 + \sin^2 \theta d\tilde \Omega_3^2 ].
\ee

\begin{figure}
        \centerline{\psfig{file=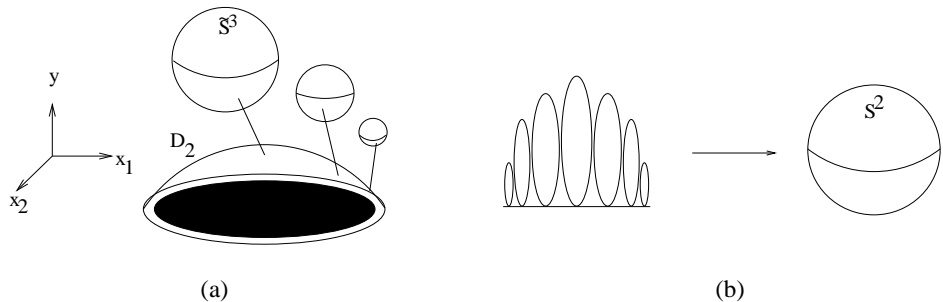,width=0.75\textwidth}}
        \caption{a) $S^5$ as a degenerate $\tilde S^3$ fibration over the disk $D_2$. b) ${ S}^2$ as a degenerate $S^1$ fibration over the interval, for comparison.}
        \label{fig:spheres}
\end{figure}

We can identify the $S^5$ from the droplet in Figure \ref{fig:blob}a by considering a disk in the 3D base space with boundary in the white region at $y=0$ (Figure~\ref{fig:spheres}).  By fibering $\tilde S^3$ over this disk, we reconstruct a non-contractible $S^5$, much as an $S^2$ is realized in toric geometry as an $S^1$ fibered over an interval (Figure~\ref{fig:spheres}b).  A simple proof shows that for a generic LLM geometry, the five-form flux threading similarly constructed five-spheres is simply given by the black area enclosed, in units of $2\pi l_p^4$.  We will later arrive at the highly non-trivial topology of black hole microstates by a similar construction.
\\

In terms of the LLM ansatz, the superstar is represented by a gray disk, 
\bea
z(r,\phi,y=0) = \half- \rho_0 \theta(r_0-r) \label{disk}
\eea  
with $r_0^2 = (1+{2\Delta \over N^2})R_{AdS}^4$  and fermion density $\rho_0 = ({2\Delta \over N^2}+1)^{-1}$.  This matches (\ref{superds}) after the change of variable~\cite{Klemm}
\bea
y &=& R_{AdS} R \cos \theta \label{coords}\\
r &=& R_{AdS}^2 \sqrt{f} \sin \theta. \nonumber
\eea
Since $z(0)\neq \pm \half$ within the disk, the geometry is singular, but the intermediate filling fraction suggests a coarse-grained gas of fermions \cite{llm}.  That this averaging is necessary to make contact with a black hole is evidence of the intimate connection between black holes and statistics suggested by the laws of black hole thermodynamics, and it was this stark demonstration of the inherently statistical nature of the black hole that was the original motivation for this work.

We can now understand the entropy of the black hole as the statistical entropy of a gas of fermions, and reinterpret the smoothness condition $\rho \in \{0,1\}$ as the vanishing of the Shannon entropy on the fermion phase space,
\bea
-\rho \ln \rho =0  &\Leftrightarrow& z(0)=\pm \half. \label{Shannon}
\eea
Turning on entropy in phase space brings about high curvatures, in turn bringing about $\alpha'$ corrections in the geometry.

In the remainder of this paper, we will test the relationship between the black hole geometry and the ensemble of fermions.  We will aim to understand how the geometries account for the microstates, why the solution of \cite{Myers} is incomplete, and test the details of Mathur and Lunin's proposal \cite{Lunin}\cite{Mathur}, roughly ``naive black hole geometry=average over microstates''.

\section{Statistics: Coarse-Graining and Microscopic Entropy}

\subsection{Microscopic Entropy}
Since we are working in a supersymmetric sector at zero temperature, we will use the micro-canonical, rather than canonical ensemble as in \cite{Buchel},\footnote{It is possible that a mixed ensemble at a fixed chemical potential, rather than charge, would be appropriate.  See \cite{OSV}.} so we are interested in the degeneracy $d_{\Delta,N}$ of $N$ fermions with excitation energy $\Delta$ above the ground state energy $N^2/2$, which can be extracted from the partition function 
\be
Z(N,q) &=& Tr(q^\Delta)\\
&=& \sum_\Delta d_{\Delta,N} q^\Delta\\
&=& \sum_{n_1=0}^\infty \sum _{n_2=n_1+1}^\infty \cdots \sum _{n_N=n_{N-1}+1}^\infty q^{-{N^2\over 2}+\sum_{i=1}^N (n_i+\half)} .
\ee
This can be evaluated explicitly in the two regimes of high and low density, $\Delta/N^2 << 1$, and $>>1$, respectively.  For the high-density regime, it is convenient to change variables to $k_i= n_i-n_{i-1}-1$ \cite{Suryanarayana}, so that the sum becomes
\be
Z(N,q) &=& \left( \prod_{i=1}^N\sum_{k_i=0}^\infty \right) q^{\sum_{i=1}^N i k_i}\\
&=& \prod_{n=1}^N (1-q^n)^{-1}.
\ee
When $\Delta/N^2 << 1$, excitations are small fluctuations near the top of a deep fermi sea, insensitive to the fact that the sea has a bottom, and therefore insensitive to $N$.  Taking $N\rightarrow \infty$, this is recognized as the Dedekind $\eta$-function, or the partition function for a 1+1 dimensional chiral boson, whose degeneracy is famously given by the Cardy formula
\bea
d_{\Delta,N} \simeq {1\over 4\sqrt{3} \Delta} e^{\pi \sqrt{2\Delta \over 3}}. \label{cardy}
\eea
For a more detailed derivation, see \cite{Suryanarayana}.

In the opposite regime of $\Delta/N^2 >> 1$, the fermions are well spaced, behaving as a classical gas of identical particles for which the exclusion principle is unimportant.  The partition function is given in terms of a single-particle partition function $Z_1(q)$ as
\be
Z(N,q) &=& q^{{-N^2 \over 2}} {(Z_1(q))^N \over N!},
\ee
where the factor $q^{-{N^2/2}}$ takes account of the shift in the Hamiltonian by the fermi sea ground state energy.  With 
\be
Z_1(q) &=& \sum_{n=0}^\infty q^{n+\half}\\
&=& {q^\half \over 1-q},
\ee
we have
\be
Z(q) &=& q^{{-N^2 \over 2}} {1 \over N! (1-q)^N}\\
 &=& {q^{{-N^2 \over 2}} \over N!} \sum_{l=0} q^l \left( \begin{array}{c}
l+N-1\\
N-1\\
\end{array}\right)\\
 &\simeq& {1 \over N!} \sum_{\Delta=0} q^\Delta \left( \begin{array}{c}
\Delta\\
N\\
\end{array}\right),
\ee
allowing us to identify $d_{\Delta,N}$ as the coefficient of $q^\Delta$,
\bea
d_{\Delta,N} &=& {\Delta! \over N!^2 (\Delta-N)!}\nonumber \\
&\simeq&  {\Delta^N \over N!^2}. \label{diluted}
\eea
We have used $(A+B)!/A! \simeq (A+B)^B \sim A^B$, for $A >> B$.  This yields an unusual form for the entropy
\bea
S_{\Delta, N} &=& \ln(d_{\Delta,N})\nonumber\\
&\simeq& N \ln {\Delta \over N^2}.\label{Slow}
\eea

\subsection{Average Distribution}
In order to test the relation between the superstar geometry (\ref{superds}) and the microstate geometries, we will find the average phase-space distribution of fermions in the micro-canonical ensemble of fixed excitation energy $\Delta$.  

Mathur and Lunin's proposal \cite{Lunin}\cite{Mathur} that black hole geometries are an average over microstate geometries generally suffers from the ambiguity of the choice of sum one would use to add geometries in the average.  In this case, however, the statistical mechanics of the dual free-fermion description provides a natural way to coarse-grain over microstate sources.

The probability that some level $n$ is occupied is the fraction of the total $d_{\Delta,N}$ states with a fermion at level $n$:
\begin{eqnarray}
P(n)_{\Delta,N}&=& {1 \over d_{\Delta,N}} \sum_{n_1=0}^\infty \sum_{n_2=n_1+1}^\infty \dots \sum_{n_N=n_{N-1}+1}^\infty \delta_{\sum_{i=1}^N n_i,\Delta+\half N(N-1)}\sum_{i=1}^N \delta_{n_i,n}. \label{prob}
\end{eqnarray}
The first $\delta$ enforces the ensemble and the second counts whether a fermion is at level $n$.  To evaluate this, we make use of a recursion relation,
\be
d_{\Delta,N} P(n)_{\Delta,N}= d_{\Delta+N-1-n,N-1}(1-P(n)_{\Delta+N-1-n,N-1}).
\ee
This can be understood as follows.  If we first imagine that the gas of fermions is quite dilute, pinning one fermion at level $n$ leaves a gas of $N-1$ fermions with excitation energy $\Delta + N -1 -n$ above an $(N-1)$-particle fermi sea.  These $N-1$ fermions have $d_{\Delta+N-1-n,N-1}$ accessible states, so
\be
P(n)_{\Delta,N} &\simeq& {d_{\Delta+N-1-n,N-1}\over d_{\Delta,N}}.
\ee
However, this is not quite right if the fermion gas is not very dilute.  Of these $d_{\Delta+N-1-n,N-1}$ states, a fraction $P(n)_{\Delta+N-1-n,N-1}$ would have a fermion at level $n$, and should not be counted.  Thus the correct form is
\be
P(n)_{\Delta,N} &=& {d_{\Delta+N-1-n,N-1}\over d_{\Delta,N}}(1-P(n)_{\Delta+N-1-n,N-1})
\ee
which is equivalent to the above recursion relation.  This heuristic argument is confirmed by careful manipulations of equation (\ref{prob}).

With this in hand, we can use the fact that at large $N$ and $\Delta$, $P(n)_{\Delta,N}$ is utterly indistinguishable from $P(n)_{\Delta+N-1-n,N-1}$, allowing us to just solve for the distribution $P(n)$ algebraically,
\bea
P(n)_{\Delta,N} &=& \left(1+ {d_{\Delta,N} \over d_{\Delta+N-1-n,N-1}}\right)^{-1}. \label{P(n)}
\eea
\\

Plugging in the expression (\ref{cardy}) for $d_{\Delta,N}$ at high density  yields
\bea
P(n) &=& 1/[1+\exp({\pi \over \sqrt{6\Delta}}(n-N))] \hspace{.5 in} {\Delta \over N^2} << 1. \label{Phigh}
\eea
This a step function $P(n) = \theta(N-n)$ smoothed out with a width $\sim \sqrt{\Delta}$.  In the opposite regime, plugging (\ref{diluted}) into (\ref{P(n)}) yields
\bea
P(n) &\simeq& {N^2\over \Delta}(1-{n \over \Delta})^N \nonumber\\
&\simeq& {N^2 \over \Delta}e^{-nN \over \Delta}  \hspace{1 in} {\Delta \over N^2} >> 1. \label{Plow}
\eea
To test that these are correct, we can check that they satisfy the two defining constraints of the ensemble, fixed number and total energy,
\be
\int_0^\infty dn P(n) &=& N\\
\int_0^\infty dn P(n) n &=& \Delta + N^2/2.
\ee
The latter provides a particularly non-trivial check; the coefficient of $\Delta$ requires the fact that $\zeta(2)=\pi^2/6$, for example.

Neither of these distributions agrees with the solid gray disk distribution that seeds the superstar (\ref{disk}).  Both fall off on a length scale that agrees with the radius $r_0$ of the gray disk, so geometries seeded by (\ref{P(n)}) will differ from (\ref{superds}) seeded by (\ref{disk}) only in the details near the singularity, but the fact that they do not agree exactly represents disagreement with Mathur and Lunin's conjecture \cite{Lunin}\cite{Mathur} that the ``naive'' classical geometry should arise as the average over the microstate geometries in the ensemble.  The appealing proposal that ``naive black hole geometry = ensemble average'' appears to need further refinement.

\section{Stretched Horizon}

A second element of Mathur and Lunin's conjecture concerns his definition of the stretched horizon.  In the absence of a causal horizon upon which to apply the Bekenstein relation, Mathur advocates that the entropy should be encoded in the area of ``the boundary of the region where the states differ'' \cite{Mathur}.  We will estimate a surface beyond which individual microstate geometries no longer differ significantly, but find that the location of this surface depends sensitively on what we take to be a significant difference, and that the area of this surface does not reproduce the microscopic entropy calculated above.

\subsection{A Subtlety}

\begin{figure}
        \centerline{\psfig{file=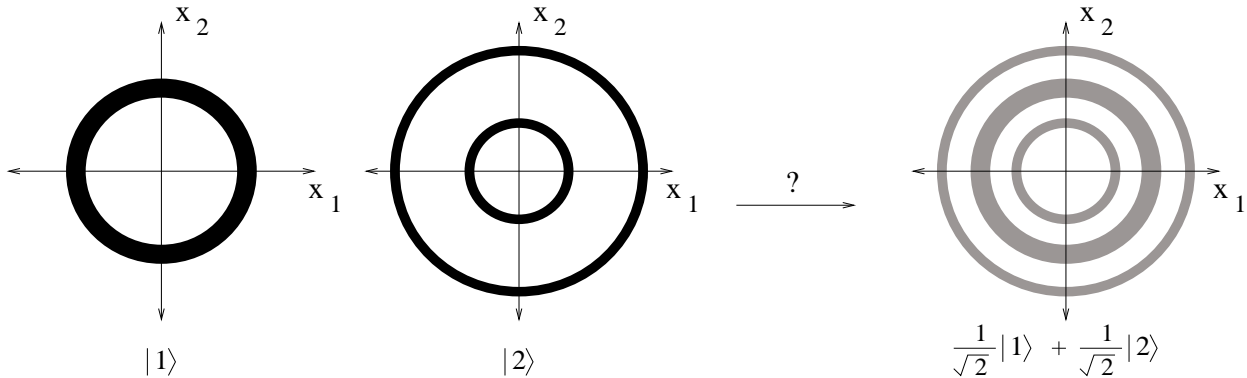,width=1\textwidth}}
        \caption{ Is the quantum linear combination of two non-singular geometries a single singular geometry?}
        \label{fig:kets}
\end{figure}

Before constructing the geometries dual to these fermionic microstates, there is an intriguing subtlety we must confront.  While the CFT description is manifestly quantum mechanical, LLM's prescription for constructing the dual geometry requires projecting to a density in a semi-classical phase space.  While it may seem natural to use a Wigner function to map an arbitrary state to a geometry, this leads to some very undesirable consequences.  For example, if states $\ket{1}$ and $\ket{2}$ each correspond to non-singular geometries, the Wigner function for the linear combination ${1 \over \sqrt{2}}\ket{1}+{1 \over \sqrt{2}}\ket{2}$ has the density $\rho=\half$, resulting in a single singular geometry, rather than a linear combination of the original geometries (Figure \ref{fig:kets}).

This prescription seems to confuse the quantum linearity of the CFT's Hilbert space with the classical linearity of the equation of motion of a function in the metric.  The problem is generic to the AdS/CFT correspondence, with a linear combination of vertex operators in the boundary giving rise to a linear combination of classical boundary conditions in the bulk \cite{KlebandWitten}.\footnote{We thank Igor Klebanov for discussion on this point.}  

Since the process of constructing a geometry is linear in the probability, not the wave function, different choices of quantum basis result in different microstate geometries.  Most choices would not lead to smooth geometries, but the requirement of smoothness is not sufficient to determine the basis.  Since our ensemble is defined by fixing the operator $\Delta$, we will choose the natural basis of eigenstates of $\Delta$.  This choice implies that every geometry in the ensemble is rotationally invariant, seeded by concentric rings in phase space, something that would not be the case with another choice.  This is to be contrasted with a choice of coherent states\footnote{Coherent states do, of course, form a vastly over-complete basis, which adds additional ambiguity to this choice.}, which may seem more natural for a classical geometry, with D-branes localized in space.  This choice of basis would yield smooth microstates that are not rotationally invariant, so the choice of basis would seem to be measurable, a puzzling state of affairs.  Clearly, there is more to be understood about the quantum mechanics of the bulk.
\\

\subsection{The Stretched Horizon}
Proceeding with the choice of eigenstates of $\Delta$, we would like to study the geometries of the ensemble.  Since (\ref{eezeeeq}) is linear, the geometry seeded by a radially-symmetric distribution of fermions $\rho(r)$ can be found as a superposition of the black disk solutions of AdS (\ref{zads}),
\be
\bz(r,y) &=& \int_0^\infty dr_0 \rho(r_0) {\partial \bzA (r,y;r_0)\over \partial r_0},
\ee
where $\bz\equiv \half-z$.  However, it is not possible to perform this integral for a generic $\rho(r)$ in the ensemble or for the average distributions (\ref{Phigh}) and (\ref{Plow}), so it is necessary to work with an approximation to the exact metrics.  With $\rho(r_0)$ localized near the origin, a multipole expansion suggests itself,
\be
\bz(r,y) &=& \sum_{l=0}^\infty f_l(r,y) \int_0^\infty dr_0 \rho(r_0) 2r_0^{2l+1},
\ee
where
\bea
f_l(r,y)&=& \left.{1\over l!}{\partial^{l+1} \bzA \over (\partial r_0^2)^{l+1}}\right|_{r_0=0}. \label{fl}
\eea
For the ensemble of fixed $N$ and $\Delta$, the first two multipoles $c_l=\int_0^\infty dr_0 \rho(r_0) 2r_0^{2l+1}$ are fixed, $c_0 = 4\pi l_p^4 N=R_{AdS}^4$ and $c_1 = 16 \pi^2 l_p^8 (\Delta +\half N^2)=R_{AdS}^8({\Delta\over N^2}+\half)$, so that differences among geometries will be governed by the multipoles $l\geq2$.  For comparison, we look at the diluted disk distribution of the superstar and at the expected-value fermion distribution, the latter in the ${\Delta \over N^2} >> 1$ regime (\ref{Plow}).  We are using the conversion between radius $r_0$ in the $x_1$-$x_2$-plane and harmonic oscillator level $n=r_0^2/4\pi l_p^4$.  We find
\be
c^{(SS)}_l &=& {2^l \over l+1}\left(R_{AdS}^4 {\Delta \over N^2}\right)^{l+1} {N^2 \over \Delta}\\
c^{(avg)}_l &=& l!\left(R_{AdS}^4 {\Delta \over N^2}\right)^{l+1} {N^2 \over \Delta} \hspace{1 in} {\Delta \over N^2} >> 1.
\ee
We see that the $l^{th}$ multipole in $r_0^2$ is governed by the distance scale $\left(R_{AdS}^4 {\Delta \over N^2}\right)^{l+1}$.  

To evaluate the average variation of $c_l$ in the ensemble, $Var(c_l)^2 = \braket{c_l^2}-\braket{c_l}^2$, we rewrite the multipoles as,
\be
c_l &=& ({R_{AdS}^4 \over N})^{l+1}\int dn n^l \rho(n)\\
&=& ({R_{AdS}^4 \over N})^{l+1} \sum_{i=1}^N n_i^l,
\ee
where we have used the definition of the fermion distribution function $\rho(n)= \sum_{i=1}^N \delta(n-n_i)$.  The variation of $c_l$ is then related to the variation of $n^l$, 
\be
Var(\sum_{i=1}^N n_i^l) &=& \sum_{i=1}^N Var (n_i^l)\\
&\simeq& N Var (n_1^l)\\
&=& N (\braket{n_1^{2l}}-\braket{n_1^l}^2)^{1/2}\\
&=& N ({1\over N}\braket{n^{2l}}-{1\over N^2}\braket{n^l}^2)^{1/2}\\
&=& (N \braket{n^{2l}}-\braket{n^l}^2)^{1/2}.
\ee
We have neglected some inter-particle correlations in order to approximate the $N$-particle distribution in terms of the one-particle distribution $\braket{n_1^l}$, which is certainly not valid for $l=0,1$, for which the ensemble enforces correlations that eliminate a variation in $c_0$, and $c_1$, but which is an excellent approximation at large $N$ for $l\geq2$.  For $l>1$ we have
\be
Var(c_l) &=& \left({R_{AdS}^4 \over N}\right)^{l+1} Var(n^l)\\
&=& \left[R_{AdS}^4\braket{c_{2l}}-\braket{c_l}^2\right]^{1/2}.
\ee
In the ${\Delta \over N^2}>>1$ regime, this becomes
\be
Var(c_l) &=& \left[R_{AdS}^4 (2l)! \left({R_{AdS}^4 \Delta \over N^2}\right)^{2l+1}{N^2 \over \Delta}-\left(l!\left({R_{AdS}^4 \Delta \over N^2}\right)^{l+1} {N^2 \over \Delta}\right)^2\right]^{1/2}\\
&=&  \left(R_{AdS}^4 {\Delta \over N^2}\right)^{l+1} {N^2 \over \Delta} \sqrt{(2l)!- l!^2}\\
&>& \braket{c_l}.
\ee
We see that the expected variation of $c_l$ is even larger than the difference between the specific $c_l$'s we saw above.
\\

We now wish to find the surface where the microstate geometries no longer differ significantly, for which we need the behavior of $f_l(r,y)$ (\ref{fl}).  The exact expressions are cumbersome, but in general they behave as
\be
f_l(r,y) &\simeq& g_l(\psi) \rho^{-2(l+1)},
\ee
where we have gone to spherical coordinates ($r=\rho \sin \psi$, $y=\rho \cos \psi$), and $g_l(\psi)$ is a $2(l+1)^{th}$ order polynomial in $\cos{\psi}$, with average $\langle g_l(\psi)\rangle= 2(-1)^l \left( \begin{array}{c} l-1\\ \lfloor {l-1 \over 2} \rfloor \end{array}\right)^2$.  Averaging over $\psi$, we have that the average fluctuation in $z$ is
\be
Var(z(\rho)) &=& \sum_{l=2}^\infty Var(c_l) \rho^{-2(l+1)} \langle g_l(\psi)\rangle\\
&=& {N^2 \over \Delta} \sum_{l=0}^\infty \hat c_l  {\left( {\Delta \over N^2}{R_{AdS}^4 \over \rho^2}\right)^{l+1}},
\ee
where $\hat c_l$ is a purely numerical coefficient.  Since $\hat c_l$ grows in $l$, the $l\geq2$ terms will significantly modify the geometry unless $\rho$ is large enough that
\be
{\Delta \over N^2 } {R_{AdS}^4 \over \rho^2} < {1 \over \Lambda},
\ee
where $\Lambda^{-1} <<1$ defines some threshold of small difference.  According to Mathur, the surface 
\be
r^2+y^2 &=& \Lambda R_{AdS}^4{ \Delta \over N^2}
\ee
is a stretched horizon, whose area should match the statistical entropy.

However, it is suspect that this surface should play the role of a horizon.  Since we are interested in a surface defined to live in a region unaffected by the choice of microstate, we are free to use the superstar metric to describe it.  In coordinates (\ref{coords}), the surface $r^2+y^2 = \Lambda R_{AdS}^4{ \Delta \over N^2}$ is $(R/R_{AdS})^2 \sim \Lambda \Delta/N^2 \gg 1$.  The function $f$ (\ref{f}) appearing in the superstar metric (\ref{superds}) demonstrates the transition from the near-star to asymptotically $AdS$ regions,
\be
f &=& 1+({R\over R_{AdS}})^2+2{\Delta \over N^2}.
\ee
Since the would-be stretched horizon is at $(R/R_{AdS})^2 >> \Delta/N^2$, the presence of the superstar demonstrated by the $2\Delta/N^2$ term is not felt, and our surface is way out in AdS, insensitive to the presence of the superstar.
\\

\subsection{Area of the Stretched Horizon}

Since this surface is defined to be in a region independent of the choice of microstate, we are free to calculate the area in the explicitly known superstar geometry.  In coordinates (\ref{coords}), the surface is at
\be
\Lambda {\Delta \over N^2} &=& {1\over R_{AdS}^4}(r^2 + y^2)\\
&=&  \left({R\over R_{AdS}}\right)^2 \cos^2 \theta + \left(1+(1+{2R_{AdS}^2\Delta \over N^2 R^2} )({R\over R_{AdS}})^2\right) \sin^2 \theta\\
&=&  \left({R\over R_{AdS}}\right)^2 + (1+{2\Delta \over N^2 }) \sin^2 \theta.
\ee
With $\Lambda>>1$ and ${\Delta \over N^2 }>>1$, the fluctuation in $\sin^2 \theta$ is insignificant, yielding 
\be
\left({R_h\over R_{AdS}}\right)^2 &\simeq& (\Lambda - 2 \sin^2 \theta){\Delta \over N^2}\\
&\simeq& (\Lambda - 1){\Delta \over N^2},
\ee
so the stretched horizon is a surface of constant $R_h=R_{AdS} {\sqrt{\Delta} \over N}\sqrt{\Lambda-1}$.  (We will take $\Lambda-1 \rightarrow \Lambda$ hereafter).

The metric (\ref{superds}) induces a metric on the $t=0$, $R_h=R_{AdS} {\Delta \over N^2}\sqrt{\Lambda}$ surface
\be
ds^2_{(8)} &=& R_{AdS}^2 ({H \over G}\sin^2 \theta d\phi^2 + G \left({R_h \over R_{Ads}}\right)^2 d\Omega_3^2\\&&+ G d\theta^2+ G^{-1}\cos ^2 \theta d \tilde \Omega_3^2)
\ee
for a volume
\bea
V_{(8)} &\sim& R_{AdS}^8 \left({R_h \over R_{AdS}}\right)^3 \sqrt{H}\nonumber\\
&\sim& R_{AdS}^8 \left({\Lambda\Delta \over N^2}\right)^{3/2}(1+{2 \over \Lambda})^{1/2}\label{vol}\\
&\sim& R_{AdS}^8 \left({\Lambda \Delta \over N^2}\right)^{3/2}.\nonumber
\eea
Up to numerical factors, we have a conjectured Bekenstein-Hawking entropy
\be
{V_{8}\over l_p^8} &\sim& {R_{AdS}^8 \Delta^{3/2} /N^3 \over R_{AdS}^8/N^2}\\
&\sim& {\Delta^{3/2} \over N},
\ee
which bears no resemblance to $S=N\ln {\Delta \over N^2}$, (\ref{Slow}).  Along the way we have made a number of somewhat arbitrary choices as to where we put the stretched horizon, focusing on the metric function $z$, for example.  In fact, repeating this calculation with other choices can yield rather dramatically different areas.  No readily apparent choice, however, reproduces the logarithmic form of (\ref{Slow}).

Repeating this analysis for the distribution in the $\Delta/N^2 <<1$ regime studied in \cite{Suryanarayana}, (\ref{Plow}) yields a horizon radius $R_h \sim R_{AdS}$, a factor $\sqrt{N}$ larger than the expression in equation (42) of \cite{Suryanarayana} needed to agree with the entropy (\ref{cardy}).  That the metric function (\ref{H})\be
H(R) &=& 1 + 2{R_{AdS}^2\over R^2}{\Delta \over N^2} \\
H(R_h) &\simeq& 1,
\ee 
takes a value independent of the superstar's charges again indicates that this surface is in a region far from the region of the superstar.\footnote{It may seem we are contradicting ourselves, saying the surface bounding the region where the geometries differ is far from the region where the superstar impacts the geometries.  We are merely stating that fluctuations between geometries effect distances far from the core of the geometries.}
\\

We see that Mathur and Lunin's definition of stretched horizon does not appear to capture the entropy of the superstar.  Indeed there are a number of ambiguities with that definition.  The first is whether we are to consider differences between {\it any} microstates, only those that represent likely fluctuations about the mean, or perhaps (as \cite{Mathur} and \cite{Suryanarayana} seem to do) the smallest such fluctuation among microstates.  The second is the proper definition of ``small''.  With large dimensionless numbers hanging around, $N$ and $\Delta$, it is unclear whether small should be defined relative to 1, $1/N$ or $1/\Delta$, say.  

We might imagine that we should be free to choose a level of accuracy, or a degree of coarse-graining, and that the entropy {\it should} vary as our threshold counts fewer or more microstates to be sufficiently similar.  However, increasing the accuracy requirement will require us to move further away, where differences are more highly suppressed, moving the stretched horizon outward, and increasing its area, as seen in (\ref{vol}), in contradiction with the fact that decreased coarse-graining should {\it lower} the entropy.

If, on the other hand, we are not free to move the threshold, but use one fixed by a comparison of fluctuations between metrics in the classical ensemble with quantum fluctuations of a single metric in the ensemble (or the averaged metric), the argument is reversed.  Intuitively, as the value of $\hbar$ is increased, quantum fluctuations, and thus the threshold for classical fluctuations are increased, causing more coarse-graining and a smaller horizon area.  This cartoon picture is in line with the behavior of the Bekenstein entropy $S\sim A/G\hbar$, which should stay fixed under an adiabatic change in $\hbar$.  Using the uncertainty principle to define what we mean by a small fluctuation also has the intriguing benefit of connecting the definition of the horizon to the value of $\hbar$, tying together its geometric and quantum roles.  

Given the problems with this definition of stretched horizon, we should ask whether the discrepancy is particularly bothersome.  In the absence of a true horizon, the ambiguity in the microstates themselves account for the entropy, and it is not clear that it ought to be encoded geometrically at all.

\section{Physical Origin of Microstates}
 
It is natural to wonder what exactly is wrong with the superstar metric (\ref{superds}) as it stands.  From the expression for the radii of the two $S^3$'s, (\ref{radii}), we see that {\it both} spheres shrink to zero size in a gray area of the $y=0$ plane, where $z(0)\neq \pm \half$.  While the $S^5$ constructed in \adsS\ by a fibration of $\tilde S^3$ could not be contracted into the the black disk region of the $y=0$ plane, a gray region presents no obstacle to continuously deforming the $S^5$ to contain more or less fermion density, so that the flux
\be
{1 \over 2 \pi^2 l_p^4}\int_{S^5} * F_5 
\ee
can be continuously varied, violating flux quantization.
\begin{figure}
        \centerline{\psfig{file=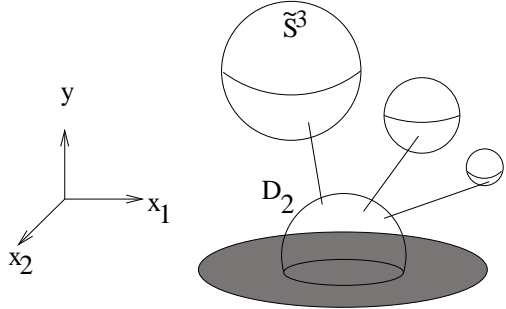,width=0.4\textwidth}}
        \caption{An example of an $S^5$ in the superstar metric which violates flux quantization.}
        \label{fig:greysphere}
\end{figure}
In order to protect quantum consistency, we must prevent the $S^5$ from contracting through regions of non-vanishing fermion density, and therefore only allow black and white regions, not gray.  Furthermore, we must give each black region an area that is an integer multiple of $2\pi l_p^4$.  This is the geometrically dual description of the requirement that fermions are discrete particles.

While the solution of Myers and Tafjord (\ref{superds}) is a good description far away from the superstar, as we approach the source RR charge, the $N$ flux lines must be parceled into integer bundles in such a way that protects flux quantization.  It is this partition of $N$ that accounts for the entropy of the black hole.  In the interior of the superstar, the partition requires a highly non-trivial topology, with a quantum foam emerging to replace the classical singularity of (\ref{superds}).\footnote{Similar behavior has been observed in \cite{Eric}, \cite{Bena}.}

\section{Conclusions and Discussion}

Our general goal has been to exploit the transparency of the LLM duality to better understand black hole entropy.  By maintaining the quantum nature of the sources, the duality demonstrates exactly how quantum effects kick in to modify the geometry away from the naive classical solution, revealing a satisfying picture: quantum mechanical flux-quantization forces us to partition the flux lines into packets of a minimal area, a process that replaces the singularity with a quantum foam.  The choice of partition directly accounts for the entropy.

In the process, we have seen vivid demonstrations of a deep connection between black holes and statistics.  An interesting feature of the result (\ref{P(n)}) for the distribution of fermions is that, since the geometries are seeded by a phase space, of all things, there is an explicit connection between the entropy and the spatial configuration of the sources of a geometry.  While the Bekenstein relation $S=A/4G$ is usually noted for its counterintuitive non-extensivity, which is taken as a cornerstone of holography, what is equally surprising is that the geometry would encode the entropy at all, that there even exists a relationship $A(S)$.  That the spatial distribution is determined by the entropy - something only clear from the dual picture - makes such a relation natural.  The fact that the geometry is seeded by a phase space, that high curvatures are triggered by Shannon entropy (\ref{Shannon}), and that the physical extent and shape of the configuration is determined by statistics offer clues to the deeply statistical nature of black holes that are not apparent from a geometric perspective alone.  The holographic point of view shows us that the geometry of the average is {\it not} an average geometry, but that averaging radically modifies the geometry in such a way that quantum and $\alpha'$ corrections are automatically turned on.  
\\

Our conclusions have offered mixed news for Mathur and Lunin's proposal.  LLM's duality provides evidence for the most controversial aspect of his proposal, that black hole geometries are the result of coarse-graining over a collection of smooth microstate geometries.  On the other hand, application of Mathur and Lunin's prescription for a stretched horizon did not succeed in reproducing the microscopic entropy, and the fact that Myers and Tafjord's solution is not seeded by the average distribution of fermions demonstrates that the prescription ``naive geometry = ensemble average'' needs to be more precisely defined, at best.  For example, it is possible that there is another ensemble whose average distribution {\it would} reproduce the homogeneous disk (\ref{disk}).
\\

Assuming that $\alpha'$ corrections to (\ref{superds}) would yield a geometry with a true horizon, it is unclear what physical interpretation to give that metric.  Is the density matrix $\rho$ somehow physical, and likewise the corresponding spacetime, or does it merely reflect an observer's lack of knowledge, and any measurement would always reveal one microstate or another?  It would be very interesting to tackle the $\alpha'$ corrections to this geometry both to test the microscopic entropy predictions and to better understand its relation to the microstates.
\\

It must be noted that it is unclear to what extent these conclusions can be generalized.  The supersymmetry and high curvatures of the superstar make this a very special case, but the mechanisms at work seem to demonstrate matters of general principle (the laws of black hole thermodynamics), suggesting that they are more general.

While it is difficult to imagine how a similar story would hold for a large Schwarzschild black hole, for which the internal density of matter can be arbitrarily small, it is especially difficult to understand how a proposal for the elimination of horizons would apply to cosmological horizons.  Every aspect of the thermodynamics of black hole horizons carries over wholesale to cosmological horizons - indeed it is possible to adiabatically swap the two for the Schwarzschild-de~Sitter solution - suggesting that a single mechanism should be behind the thermodynamics of each.  While it is plausible to imagine local corrections modifying the geometry within a black hole to eliminate its horizon, it is far more difficult to understand a mechanism that would eliminate a distant cosmological horizon.

This author remains agnostic as to how much should be inferred from the existence of microstate geometries for certain cases of black holes.

While this paper was being finalized, we were informed of related work in progress  by Balasubramanian, de~Boer, Jejjala and Sim\'{o}n \cite{Vijay2}, to appear.  For a preliminary account, see \cite{Vijay1}.

\section{Acknowledgments}
We wish to thank Andy Charman, James Gill, Eric Gimon, Igor Klebanov and Samir Mathur for useful discussions, and Shannon McCurdy for editing the draft.  We are especially grateful to Petr Ho\v rava for discussions and comments on the draft.  The results reported here were presented in June 2005 at TASI.  We are grateful to the hospitality we received in Boulder.  This material is based upon work supported in part by NSF grant PHY-0244900, by the Berkeley Center for Theoretical Physics, and by DOE grant DE-AC02-05CH11231.  This paper is dedicated to my Dad, born quite a few years ago today.  Happy Birthday, pops.

\bibliographystyle{board}
\bibliography{all}

\begin{thebibliography}{10}

\bibitem{Myers}
  R.~C.~Myers and O.~Tafjord,
  ``Superstars and giant gravitons,''
  JHEP {\bf 0111}, 009 (2001)
  [arXiv:hep-th/0109127].

\bibitem{llm}
  H.~Lin, O.~Lunin and J.~Maldacena,
  ``Bubbling AdS space and 1/2 BPS geometries,''
  JHEP {\bf 0410}, 025 (2004)
  [arXiv:hep-th/0409174].
  
\bibitem{Lunin}
  O.~Lunin and S.~D.~Mathur,
  ``Statistical interpretation of Bekenstein entropy for systems with a
  stretched horizon,''
  Phys.\ Rev.\ Lett.\  {\bf 88}, 211303 (2002)
  [arXiv:hep-th/0202072].

\bibitem{Mathur}
  S.~D.~Mathur,
  ``The fuzzball proposal for black holes: An elementary review,''
  [arXiv:hep-th/0502050].

\bibitem{AdSCFT}
 O.~Aharony, S.~S.~Gubser, J.~M.~Maldacena, H.~Ooguri and Y.~Oz,
  ``Large N field theories, string theory and gravity,''
  Phys.\ Rept.\  {\bf 323}, 183 (2000)
  [arXiv:hep-th/9905111].

\bibitem{Klemm}
  M.~M.~Caldarelli, D.~Klemm and P.~J.~Silva,
  ``Chronology protection in anti-de Sitter,''
  [arXiv:hep-th/0411203].

\bibitem{Milanesi:2005tp}
  G.~Milanesi and M.~O'Loughlin,
  ``Singularities and closed time-like curves in type IIB 1/2 BPS geometries,''
  [arXiv:hep-th/0507056].

\bibitem{us}
P.~Ho\v rava and P.~G.~Shepard,
  ``Topology changing transitions in bubbling geometries,''
  JHEP {\bf 0502}, 063 (2005)
  [arXiv:hep-th/0502127].

\bibitem{Strominger}
  A.~Strominger and C.~Vafa,
  ``Microscopic Origin of the Bekenstein-Hawking Entropy,''
  Phys.\ Lett.\ B {\bf 379}, 99 (1996)
  [arXiv:hep-th/9601029].

\bibitem{Gubser}
  S.~S.~Gubser,
  ``Curvature singularities: The good, the bad, and the naked,''
  Adv.\ Theor.\ Math.\ Phys.\  {\bf 4}, 679 (2002)
  [arXiv:hep-th/0002160].

\bibitem{Dabholkar}
  A.~Dabholkar,
  ``Exact counting of black hole microstates,''
  arXiv:hep-th/0409148.


\bibitem{Jev}
  S.~Corley, A.~Jevicki and S.~Ramgoolam,
  ``Exact correlators of giant gravitons from dual N = 4 SYM theory,''
  Adv.\ Theor.\ Math.\ Phys.\  {\bf 5}, 809 (2002)
  [arXiv:hep-th/0111222].

\bibitem{Beren}
  D.~Berenstein,
  ``A toy model for the AdS/CFT correspondence,''
  JHEP {\bf 0407}, 018 (2004)
  [arXiv:hep-th/0403110].

\bibitem{Kleb}
  I.~Klebanov,
  ``String Theory in Two Dimensions,''
  [arXiv:hep-th/9108019].

\bibitem{Jab}
M.~M.~Sheikh-Jabbari and M.~Torabian,
``Classification of all 1/2 BPS solutions of the tiny graviton matrix theory,''
[arXiv:hep-th/0501001].
\hfill \break
M.~M.~Sheikh-Jabbari,
``Tiny graviton matrix theory: DLCQ of IIB plane-wave string theory, a
conjecture,''
JHEP {\bf 0409}, 017 (2004)
[arXiv:hep-th/0406214].

\bibitem{Buchel}
  A.~Buchel,
  ``Coarse-graining 1/2 BPS geometries of type IIB supergravity,''
  arXiv:hep-th/0409271.

\bibitem{OSV}
  H.~Ooguri, A.~Strominger and C.~Vafa,
  ``Black hole attractors and the topological string,''
  Phys.\ Rev.\ D {\bf 70}, 106007 (2004)
  [arXiv:hep-th/0405146].

\bibitem{Suryanarayana}
  N.~V.~Suryanarayana,
  ``Half-BPS giants, free fermions and microstates of superstars,''
  [arXiv:hep-th/0411145].

\bibitem{KlebandWitten}
  I.~R.~Klebanov and E.~Witten,
  `AdS/CFT correspondence and symmetry breaking,''
  Nucl.\ Phys.\ B {\bf 556}, 89 (1999)
  [arXiv:hep-th/9905104].

\bibitem{Eric}
  P.~Berglund, E.~G.~Gimon and T.~S.~Levi,
  ``Supergravity microstates for BPS black holes and black rings,''
  arXiv:hep-th/0505167.

\bibitem{Bena}
  I.~Bena and N.~P.~Warner,
  ``Bubbling supertubes and foaming black holes,''
  [arXiv:hep-th/0505166].

\bibitem{Vijay2}
  V.~Balasubramanian, J.~de~Boer, V.~Jejjala and J.~Simon,
  ``The library of Babel: On the Origin of Gravitational Thermodynamics,'' to appear

\bibitem{Vijay1}
  V.~Balasubramanian, V.~Jejjala and J.~Simon,
  ``The library of Babel,''
  [arXiv:hep-th/0505123].

\end{thebibliography}
\end{document}